\documentclass[twocolumn,showpacs,preprintnumbers,prd,superscriptaddress,nofootinbib]{revtex4-1}
\usepackage{graphicx}
\usepackage{epsf}
\usepackage{bm}
\usepackage{amsmath}
\usepackage{amsfonts}
\usepackage{amssymb}
\usepackage{epstopdf}
\usepackage{natbib}
\usepackage{hyperref}
\usepackage{color}
\usepackage{verbatim}
\usepackage{multirow}
\usepackage{bm}
\usepackage{hyperref}
\usepackage{float}
\usepackage{xcolor}

\definecolor{darkblue}{rgb}{0.0, 0.0, 0.55}
\definecolor{darkred}{rgb}{0.55, 0.0, 0.0}

\usepackage{hyperref}
\hypersetup{
    colorlinks=true, 
    linkcolor=darkblue,
    citecolor=darkblue,
    urlcolor=darkblue}
    
\makeatletter\let\expandableinput\@@input\makeatother

\begin{document}

\title{Constraining Cosmological and Astrophysical Parameters with the Cosmic
Star Formation History}

\author{Miguel Moyses}
\email{miguel.moyses@ufrgs.br}
\affiliation{Instituto de F\'{i}sica, Universidade Federal do Rio Grande do Sul, 91501-970 Porto Alegre RS, Brazil}

\author{Rafael C. Nunes}
\email{rafadcnunes@gmail.com}
\affiliation{Instituto de F\'{i}sica, Universidade Federal do Rio Grande do Sul, 91501-970 Porto Alegre RS, Brazil}
\affiliation{Divisão de Astrofísica, Instituto Nacional de Pesquisas Espaciais, Avenida dos Astronautas 1758, São José dos Campos, 12227-010, São Paulo, Brazil}

\begin{abstract}
Identifying new observational probes to constrain cosmological parameters has become an important goal in modern cosmology. In this work, we explore the potential of the cosmic star formation rate density (SFRD), compiled over the redshift range $z \in [0, 15]$, as a complementary probe of fundamental parameters, including $\Omega_{\rm m}$, $H_0$, and the dark energy equation-of-state parameter, $w$. Within the $\Lambda$CDM framework, SFRD combined with BBN data alone yields $H_0 = 65\pm11$ km\,s$^{-1}$\,Mpc$^{-1}$, reflecting significant degeneracies with astrophysical parameters. By jointly analyzing SFRD with recent BAO and Type Ia supernova (SNIa) data, these degeneracies are effectively broken, resulting in much tighter constraints, e.g., \texttt{SFRD + BBN} + \texttt{DESI-DR2} gives $H_0 = 68.28 \pm 0.18$ km\,s$^{-1}$\,Mpc$^{-1}$. We perform a statistical reconstruction of the SFRD as a function of redshift, finding a peak at $z_{\rm peak} = 2.600^{+0.114}_{-0.087}$ within $\Lambda$CDM. Our results demonstrate that combining SFRD with established cosmological probes not only improves constraints on cosmological parameters but also reduces uncertainties in astrophysical parameters governing star formation. We further extend the analysis to the $w$CDM model, highlighting the promise of SFRD as a robust complementary cosmological probe across different dark energy scenarios.

\end{abstract}

\maketitle

\section{Introduction}
The study of galaxies and their properties, such as the star formation rate (SFR) and luminosity—is fundamental, as their behavior depends on both environment and cosmic epoch \cite{2023ApJS..265....5H, 2004MNRAS.351.1151B, Dav__2017, Lara_L_pez_2010, Mateus_2007, Kelkar_2017, 2025arXiv250912922K, robertsborsani2026jwstspectroscopicinsightsdiversity,  Abel:2001pr, 2014ApJS..214...15S, 2006ApJ...647..201C, 2018MNRAS.480.4379C}. Galaxy luminosity is observed to correlate linearly with the SFR, indicating that the emitted light predominantly arises from their stellar populations \cite{2024Galax..12...37B, 2005MNRAS.358.1231L, 2010ApJ...714.1256C, 2019MNRAS.482..560M, 10.1093/mnras/staa2561, 2021A&A...653A..36M}. Measuring the SFR thus provides insight into galaxy evolution, since star formation is a primary mechanism of energy production in the Universe. The radiation we observe from galaxies allows us to trace their assembly history, probe the large-scale structure, and investigate the role of dark matter \cite{Madau:2014bja, 2013MNRAS.428.3121M, Perez-Gonzalez:2025bqr}.

The most direct way to measure the SFR is by counting young stars or recent star-forming events, such as supernovae \cite{Botticella_2012, Jiao_2025}. However, this method is limited to the Milky Way and nearby galaxies. For most galaxies, the SFR is instead inferred from their integrated light, assuming a given initial mass function (IMF) and stellar formation history(SFH).  Since the observed light includes contributions from both young and old stars, specific spectral ranges are used to isolate recent star formation. Ultraviolet (UV) wavelengths are commonly employed, as they are dominated by emission from young, massive stars and thus provide a direct tracer of ongoing star formation \cite{Madau:2014bja, 10.1111/j.1365-2966.2008.12866.x, 2025ApJ...994...14P, 2026MNRAS.545f1995B, 2012MNRAS.421...98D}.

While the SFR of individual galaxies is a primary focus in astrophysics, cosmology aims to understand the Universe on the largest scales. In this context, the star formation rate density (SFRD), $\dot{\rho}_\star$, is a convenient quantity, representing the amount of star formation per unit time and comoving volume. Semi-analytic and hydrodynamical models can predict the SFRD; notably, Hernquist and Springel \cite{Hernquist:2002rg} derived its redshift evolution using a physically motivated hydrodynamical model. Several other approximations from different models were also discussed in the literature \cite{Vogelsberger_2013, Fakhry:2025yeu, Ascasibar:2002qy, Vogelsberger:2014dza, 2017ARA&A..55...59N}.

Over the past two decades, deep photometric and spectroscopic surveys have transformed our understanding of galaxy populations across cosmic time, enabling increasingly precise reconstructions of the cosmic star formation history. Observations with the Hubble Space Telescope (HST) and, more recently, the James Webb Space Telescope (JWST) have extended SFRD estimates to higher redshifts. For example, \cite{2015ApJ...810...71F} analyzed the ultraviolet luminosity function (UVLF) up to $z\sim8$, providing key constraints on the SFRD in young galaxies. \cite{Perez-Gonzalez:2025bqr}, using JWST MIDIS+NGDEEP data, have pushed this frontier to $z\sim17$--$25$, suggesting that early star formation may have been more abundant than previously anticipated. These analyses typically derive the UVLF and convert it to SFRD estimates, assuming an IMF and applying dust attenuation corrections. See \cite{2026MNRAS.546f2267C} for other recent measurements.

In this work, we propose using the SFRD as a complementary cosmological probe, primarily aimed at constraining the astrophysical parameters of the adopted model within the $\Lambda$CDM and $w$CDM frameworks, while simultaneously fitting the main cosmological parameters of each model. The SFRD acts as a background probe, sensitive to parameters such as $\Omega_m$, $H_0$, and the dark energy equation of state parameter, $w$. Moreover, we find that combining SFRD measurements with other key cosmological observables, such as baryon acoustic oscillations (BAO) and type Ia supernovae (SNIa), significantly improves both the constraints on, and the accuracy of, astrophysical parameters related to star formation. Therefore, the joint analysis of SFRD measurements and external cosmological data enables us to reconstruct the full parameter space of the SFRD with high precision.

This paper is organized as follows. In Section~\ref{model}, we describe the theoretical framework and methodology adopted to model the cosmic SFRD. Section~\ref{data} presents the datasets and the statistical methods used in this analysis. In Section~\ref{results}, we discuss our main findings. Finally, in Section~\ref{final}, we summarize our conclusions and outline prospects for future research at the interface between SFRD and cosmological modeling.

\section{Modeling the Cosmic Star Formation Rate Density}
\label{model}

In this work, we adopt the minimal semi-analytical framework proposed by \cite{Hernquist:2002rg} to model the cosmic star formation rate density (SFRD). This physically motivated approach provides a functional description of the SFRD as a function of redshift, derived from the underlying processes governing gas cooling, halo collapse, and star formation within dark matter structures. The framework offers a transparent interpretation of the physical mechanisms driving the evolution of the cosmic star formation history, while simultaneously furnishing theoretical support for the empirical fitting functions commonly used to describe it. Identifying a universal behaviour from a physically grounded model is of particular interest for cosmology, as it enables robust predictions for the star formation history in cold dark matter universes. These predictions can then be directly confronted with observations to test the validity of the hierarchical galaxy formation paradigm. In the following, we summarise the key features of this theoretical model.

The model distinguishes between two principal regimes for the star formation rate: one dominating at high redshift and another at low redshift. At late times, the evolution of the SFRD, $\dot{\rho}_\star$, is primarily regulated by the declining efficiency of gas cooling, which is itself controlled by the decreasing mean density of the Universe. This physical dependence leads to a scaling with the cosmic expansion rate, expressed in terms of the Hubble parameter, $H(z)$. Empirically, the low-redshift behaviour can be approximated by
\begin{equation}
\dot{\rho}_\star \propto H(z)^{4/3}.
\end{equation}
At earlier epochs, the SFRD instead exhibits a nearly exponential decay,
\begin{equation}
\dot{\rho}_\star \propto \exp(-z/3),
\end{equation}
which likely reflects the evolution of the halo mass function (HMF), whose high-mass tail is exponentially suppressed at increasing redshift.

The general expression derived by \cite{Hernquist:2002rg} encapsulates both regimes:
\begin{equation}
\label{main_SFR}
\dot{\rho}_\star(z) = \dot{\rho}_{\star,0} \,
\frac{\chi^2}{1 + \alpha (\chi - 1)^3 \exp(\beta \chi^{7/4})},
\end{equation}
where $\dot{\rho}_{\star,0}$, $\alpha$, and $\beta$ are free parameters, and
\begin{equation}
\label{Hz_main}
\chi = \left( \frac{H(z)}{H_0} \right)^{2/3}.
\end{equation}
Equation~\eqref{main_SFR} represents the theoretical prescription we employ to model the redshift evolution of the SFRD.

Physically, the model assumes that star formation occurs exclusively within virialised dark matter haloes, allowing the cosmic SFRD to be expressed as an integral over the halo mass function. In this picture,
\begin{equation}
\dot{\rho}_\star(z) = \bar{\rho}\int g(M,z)\, s(M,z)\, \mathrm{d}logM,
\end{equation}
where $\bar{\rho}$ = $3\Omega_mH_0^2$/$(8\pi G)$, $g(M,z)$ denotes the halo multiplicity function, and 
$s(M,z) = \langle \dot{M}_\star \rangle / M$ is the mass-normalised average star formation rate. Haloes are defined as spherical overdensities enclosing 200 times the critical density of the Universe, $\rho_{\mathrm{c}}(z)$, in agreement with numerical simulations that identify this threshold with virialised structures. The condition $\rho(<R_{\mathrm{vir}}) = 200\, \rho_{\mathrm{c}}$ implies that haloes are strongly overdense relative to the cosmic mean, consistent with their nonlinear collapse.

The astrophysical factor $s(M,z)$ is obtained from hydrodynamical simulations and depends primarily on the virial temperature, $T_{\mathrm{vir}}$. Star formation is only efficient in haloes with $T_{\mathrm{vir}} \gtrsim 10^4\,\mathrm{K}$, above which atomic cooling becomes effective. For $10^4\,\mathrm{K} \lesssim T_{\mathrm{vir}} \lesssim 10^6\,\mathrm{K}$, $s(M,z)$ remains approximately constant, while for more massive haloes ($T_{\mathrm{vir}} \gtrsim 10^6\,\mathrm{K}$) star formation becomes increasingly efficient, as stellar feedback is less able to drive gas outflows from deeper potential wells. Although $s(M,z)$ exhibits some mass dependence, its shape is largely preserved across cosmic time, varying mainly in amplitude with redshift. This behaviour underpins the robustness of the Hernquist–Springel formalism and its applicability to a broad range of cosmological conditions.

In Eq.~(\ref{main_SFR}), $\dot{\rho}_{\star,0} \equiv \dot{\rho}_\star(z=0)$ denotes the present-day star formation rate per comoving volume. The parameter $\alpha$ regulates the early-time rise of star formation by controlling how rapidly halos above the cooling threshold assemble. The parameter $\beta$ governs the high-redshift suppression and depends on the matter power spectrum through the variance $\sigma_4^2$ at the mass scale corresponding to $T \sim 10^4\,\mathrm{K}$. It can be expressed as
\begin{equation}
\beta = \frac{a\,\delta_{\rm c}^2}{2\,\sigma_4^2(z=0)},
\end{equation}
where $a$ is the Sheth--Tormen parameter \citep{Sheth:2001dp} \footnote{The halo mass function beyond the $\Lambda$CDM framework has been investigated in a variety of scenarios (see, for example, \cite{Bhattacharya:2010wy,Li:2024wco,Shen:2024cio,Castro:2025tel}). Nevertheless, as discussed in \cite{Li:2024wco}, even for relatively general and degenerate dynamical dark energy models, such as the $w_0w_a$CDM parametrization, the Sheth--Tormen formalism retains an accuracy at the $\sim 5\%$ level without significant recalibration. This suggests that, for simpler extensions such as $w$CDM models, the approximation remains robust at a comparable or better level of precision (i.e., $\lesssim 5\%$), although small systematic deviations cannot be entirely ruled out.}
and $\delta_{\rm c}$ is the critical overdensity for collapse. Hence, $\beta$ provides a direct link between the cosmic star formation history and the underlying cosmology.

Let us now interpret the main astrophysical parameters appearing in Eq.~(\ref{main_SFR}), i.e, $\alpha$ and $\beta$.
The parameter $\alpha$ controls the strength of the suppression mechanisms regulating star formation at high redshifts. Larger values of $\alpha$ enhance the suppression term in the denominator of the SFRD, leading to an earlier saturation of the star formation rate, while smaller values allow for a more extended star formation history over cosmic time.
From a physical perspective, $\alpha$ should be interpreted as an effective phenomenological parameter. Although it does not correspond to a single physical process, it encapsulates the combined impact of several astrophysical mechanisms, such as feedback processes (e.g., supernovae and AGN activity), gas cooling efficiency, and the efficiency of halo formation. Therefore, $\alpha$ acts as a proxy for the overall regulation of star formation rather than a direct measure of a specific physical effect.

In contrast, the parameter $\beta$ is primarily associated with gravitational structure formation. It is related to the abundance of dark matter halos with virial temperatures $T_{\rm vir} > 10^4\,\mathrm{K}$, which are able to efficiently cool and form stars. In this sense, $\beta$ quantifies the suppression of star formation at high redshifts due to the scarcity of such halos in the early universe, and is therefore more directly connected to large-scale structure formation than to the local astrophysical processes described by $\alpha$.

It is important to emphasize that $\alpha$ and $\beta$ are effective parameters that encapsulate a variety of complex and interrelated astrophysical processes. Rather than attempting to model these processes explicitly---for example, including $\mathrm{H}_2$ cooling, which may play a significant role in low-mass halos at high redshift \citet{Nadler:2025mnz}, or the impact of the halo mass function on star formation \citet{Reischke:2015jga}---we treat $\alpha$ and $\beta$ as free parameters in our main analysis, to be directly sampled and constrained by the data. This strategy allows unmodeled or poorly understood astrophysical effects to be effectively absorbed into their inferred values through the global fit. Consequently, $\alpha$ and $\beta$ should be interpreted as effective global fitting parameters, consistent with the general interpretation outlined above.

\begin{figure*}
    \centering
    \includegraphics[width=0.45\textwidth]{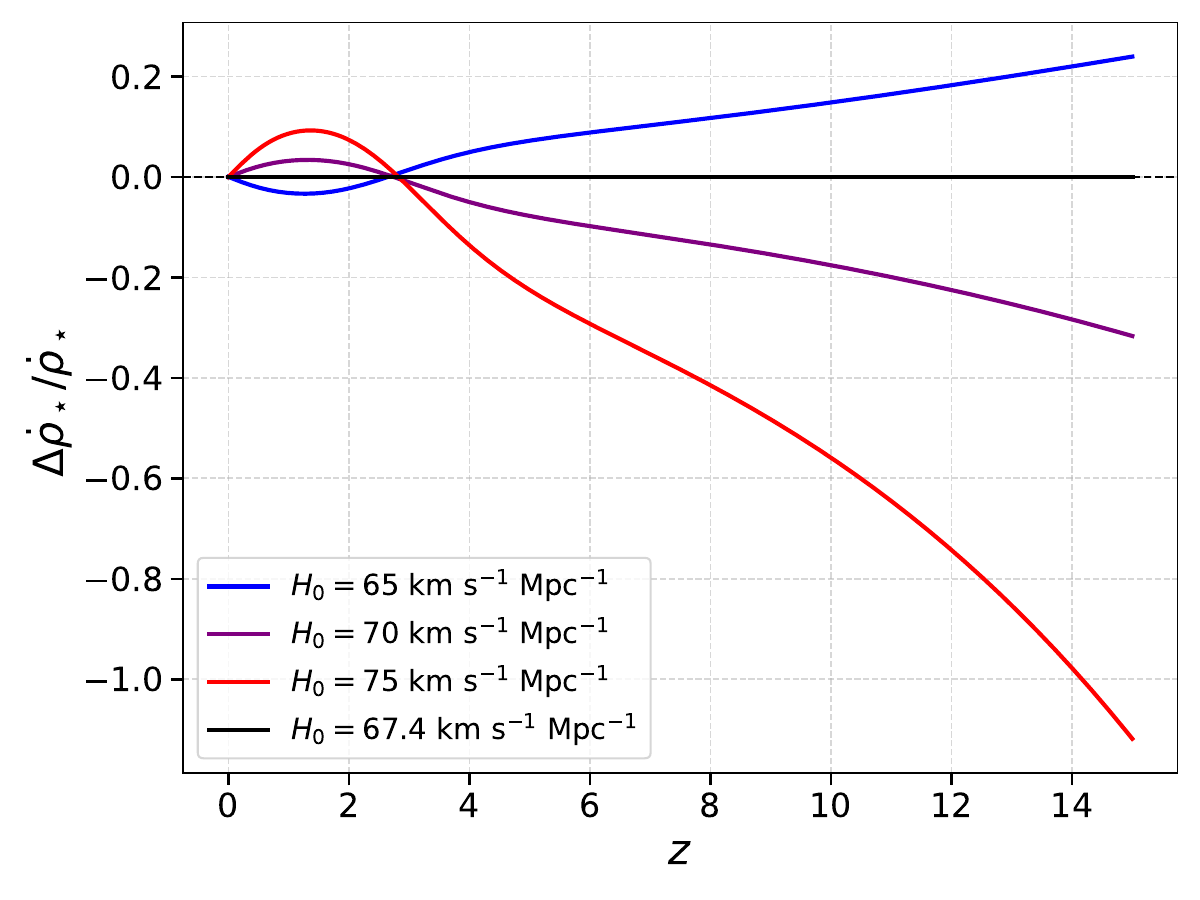}
    \hspace{0.04\textwidth}
    \includegraphics[width=0.45\textwidth]{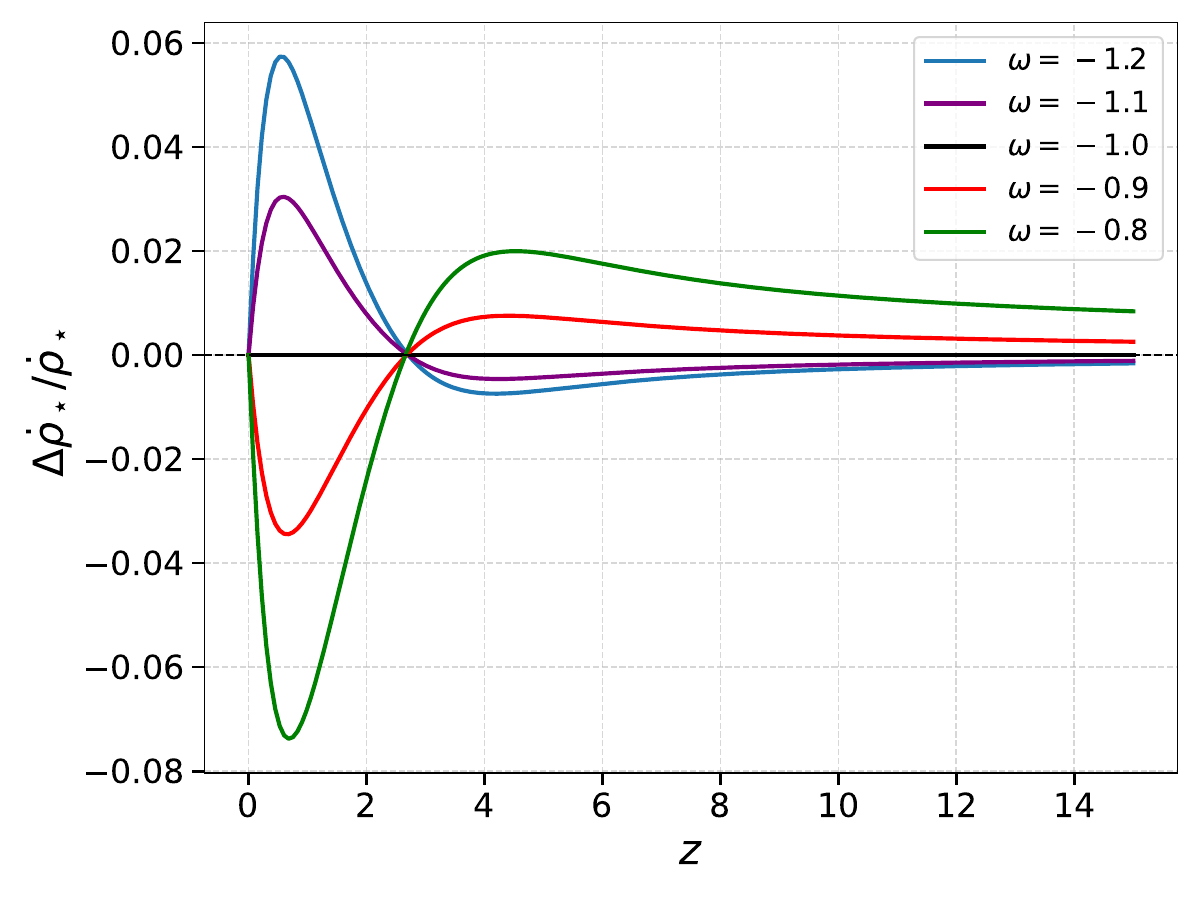}
    \caption{
        \textit{Left:} Theoretical relative difference in the SFRD, given by Eq.~(\ref{Diff_curvas}), for different values of the Hubble constant $H_0$. 
        \textit{Right:} Same as the left panel, but for variations in the dark energy equation-of-state parameter $w$.
    }
    \label{fig:Diff_curvas}
\end{figure*}

The large-scale structure of the Universe is accurately described by the Friedmann–Lemaître–Robertson–Walker (FLRW) metric, which assumes that spacetime is both homogeneous and isotropic. Under the assumption that General Relativity provides a valid description of the dynamics of cosmic expansion, the evolution of the Hubble parameter, $H(z) \equiv \dot{a}/a$, in Eq.~\eqref{Hz_main}, is governed by the Friedmann equation, which can be written as
\begin{align}
\frac{H(z)}{H_0} &=
\Bigg[ (\Omega_{b} + \Omega_{\mathrm{cdm}})(1 + z)^3 +
\Omega_{\gamma}(1 + z)^4 + \Omega_{K}(1 + z)^2 \nonumber\\
&\quad + \Omega_{\nu} \frac{\rho_{\nu}(z)}{\rho_{\nu,0}} +
\Omega_{\mathrm{DE}} \frac{\rho_{\mathrm{DE}}(z)}{\rho_{\mathrm{DE},0}} \Bigg]^{1/2} \,.
\label{eq:Friedmann}
\end{align}
Here, $\Omega_b$, $\Omega_{\mathrm{cdm}}$, $\Omega_{\gamma}$, $\Omega_{K}$, $\Omega_{\nu}$, and $\Omega_{\mathrm{DE}}$ denote the present-day fractional energy densities of baryons, cold dark matter, photons, curvature, neutrinos, and dark energy, respectively. In this work, we assume a spatially flat FLRW universe ($\Omega_k = 0$). Neutrinos are modeled under the normal mass hierarchy, with a total mass fixed to $\sum m_{\nu} = 0.06~\mathrm{eV}$. These assumptions are consistent with the standard baseline adopted in cosmological analyses, facilitating comparison with the literature while minimizing the dimensionality of the parameter space. All results presented here are obtained within this fiducial framework, unless otherwise specified.

Within the standard $\Lambda$CDM framework, dark energy is modeled as a cosmological constant, $\Lambda$, characterized by an energy density $\rho_{\mathrm{DE}}$ that remains constant in both space and time. More generally, dark energy may be described by an equation-of-state parameter
\begin{equation}
w(z) \equiv \frac{P(z)}{\rho_{\mathrm{DE}}(z)} \,,
\end{equation}
where $P(z)$ is its pressure. The corresponding evolution of the dark energy density is then given by
\begin{equation}
\frac{\rho_{\mathrm{DE}}(z)}{\rho_{\mathrm{DE},0}} =
\exp \left[ 3 \int_0^z \frac{1 + w(z')}{1 + z'} \, dz' \right] \,.
\label{eq:rhoDE}
\end{equation}

For a constant equation-of-state parameter $w$, Eq.~\eqref{eq:rhoDE} simplifies to
$\rho_{\mathrm{DE}}(z)/\rho_{\mathrm{DE},0} = (1 + z)^{3(1 + w)}$, with $w = -1$ corresponding to a true cosmological constant.

Figure~\ref{fig:Diff_curvas} shows the theoretical relative difference in the SFRD, quantified as

\begin{equation}
\label{Diff_curvas}
\frac{\Delta \dot{\rho}^{}_\star(z)}{\dot{\rho}^{}_\star(z)} = \frac{\dot{\rho}^{\rm Planck}_\star(z) - \dot{\rho}_\star(\theta_i, z)}{\dot{\rho}^{\rm Planck}_\star(z)},
\end{equation}
where $\dot{\rho}^{\rm Planck}_\star(z)$ denotes the SFRD predicted using the cosmological parameters fixed at the \textit{Planck} 2018 best-fit values, combined with the astrophysical parameters obtained from the \texttt{Base} + \texttt{DESI-DR2} best-fit results (to be discussed later). The vector $\theta_i$ represents variations in specific cosmological parameters, as indicated in the legend of Figure~\ref{fig:Diff_curvas}.

In the left panel of Figure~\ref{fig:Diff_curvas}, we observe interesting trends that deviate from the theoretical expectation for the behavior of the SFRD as a function of $z$. Small deviations in the Hubble constant $H_0$ can lead to significant changes in $\dot{\rho}_\star(z)$. For redshifts $z < 2$, variations in $H_0$ produce only modest changes, typically at the level of $\sim5\%$. However, at higher redshifts, these deviations become increasingly relevant — for instance, for $H_0 = 75~\mathrm{km\,s^{-1}\,Mpc^{-1}}$, the relative difference can reach up to $\sim100\%$ at $z \sim 14$. In general, larger values of $H_0$ (above $70~\mathrm{km\,s^{-1}\,Mpc^{-1}}$) tend to predict systematically higher SFRDs at early times. This behaviour indicates that SFRD data at high redshifts ($z \gtrsim 5$) could become a valuable probe of $H_0$, with the potential to constrain or rule out higher values. 

The right panel of Figure~\ref{fig:Diff_curvas} explores the impact of varying the dark energy equation-of-state parameter $w$ on the SFRD. As expected, the effects of $w$ are less pronounced than those of $H_0$. Deviations from $w = -1$ mainly affect $\dot{\rho}_\star(z)$ at low redshifts ($z \lesssim 2$), where the cosmic dynamics are dominated by dark energy and the Universe experiences accelerated expansion. For reasonable variations in $w$, as shown in the figure, the corresponding impact on the SFRD remains below $\sim8\%$.

In what follows, all statistical inferences are performed within the context of the $\Lambda$CDM and $w$CDM models, i.e. assuming a constant dark energy equation-of-state parameter $w = \mathrm{const}$.

\section{Datasets and methodology}
\label{data}
To test the theoretical structures explored in this work, we implemented our model in the \texttt{CLASS} Boltzmann solver~\cite{Blas:2011rf} and used the \texttt{MontePython} sampler~\cite{Brinckmann:2018cvx, Audren:2012wb} to perform Monte Carlo analyses via Markov Chains (MCMC). 

The cosmological parameters sampled in this work include: the physical baryon density $\omega_{\rm b} = \Omega_{\rm b} h^2$, the physical dark matter density $\omega_{\rm c} = \Omega_{\rm c} h^2$ and the Hubble constant $H_{0}$. In addition to these standard parameters, we also sample $w$ within the context of $w$CDM models.

In all our joint analyses, convergence was verified using the Gelman–Rubin diagnostic~\cite{Gelman_1992}, requiring that \( R - 1 \leq 10^{-2} \) for all parameter chains. The statistical analysis was carried out using the \texttt{GetDist} package,\footnote{\url{https://github.com/cmbant/getdist}}, which enabled the extraction of numerical results, including one-dimensional posterior distributions and two-dimensional marginalized probability contours. In the following sections, we describe the likelihood functions and the analysis methodology adopted throughout this work.
\\
 
\begin{itemize}
\item \textit{Star Formation Rate Density} (\textbf{SFRD}): The cosmic star formation rate density (SFRD) traces the rate at which baryonic matter is converted into stars across cosmic time, providing a direct observational link between galaxy evolution and the growth of large-scale structure. Estimates of the SFRD are primarily derived from rest-frame ultraviolet (UV) luminosity functions (LFs), which probe the light emitted by young, massive stars and therefore serve as a sensitive indicator of ongoing star formation activity.  

At intermediate redshifts, we include results from \cite{2015ApJ...810...71F}, based on \textit{HST} and \textit{Spitzer} observations covering $4 \leq z \leq 8$, together with the widely used compilation of \cite{Madau:2014bja}, which combines multiple ground- and space-based surveys to trace SFRD evolution from $z \sim 0$ to $z \sim 8$. At higher redshifts, we incorporate recent constraints from \textit{JWST} observations that have significantly extended the empirical frontier of star formation studies: 
\cite{harikane2023purespectroscopicconstraintsuv} using NIRSpec spectroscopy ($8.6 \leq z_{\mathrm{spec}} \leq 13.2$); 
\cite{2024MNRAS.533.3222D} combining NIRCam imaging from the PRIMER, JADES, and NGDEEP programmes ($8.5 < z < 15.5$); 
\cite{finkelstein2023completeceersearlyuniverse} based on CEERS NIRCam imaging ($8.5 \leq z \leq 14.5$); 
\cite{2024MNRAS.527.5004M} compiling 13 wide-area \textit{JWST} surveys ($9.5 < z < 12.5$); 
\cite{willott2024steepdeclinegalaxyspace} from the CANUCS survey ($8 \leq z \leq 12$); and 
\cite{Bouwens_2023}, who employ the SMACS0723, GLASS, and CEERS fields to extend constraints up to $z \sim 17$.  

The methodologies used to infer UV luminosity functions vary across these works — including photometric selection, spectroscopic confirmation, and multi-field stacking analyses — yet they provide mutually consistent SFRD estimates within their respective uncertainties. To ensure uniformity, we prioritise datasets that assume the same stellar initial mass function (IMF; e.g. \citealt{1955ApJ...121..161S}), thereby guaranteeing a homogeneous conversion from UV luminosity to star formation rate. Residual differences in survey volume estimation remain unavoidable: while some analyses employ the maximum-volume ($V_{\mathrm{max}}$) method, others derive effective comoving volumes accounting for survey geometry, completeness, and detection efficiency. Nevertheless, these methodological variations yield results consistent within the quoted statistical and systematic uncertainties.  

Taken together, these measurements, comprising 32 individual SFRD determinations spanning $0 < z \lesssim 15$, form a comprehensive and continuous observational reconstruction of the cosmic star formation history. This unified compilation is hereafter referred to as \texttt{SFRD}.
\\

\item \textit{Baryon Acoustic Oscillations} (\textbf{DESI-DR2}): We use BAO measurements from the second data release of the DESI survey, which includes data from galaxies, quasars~\cite{DESI:2025zgx}, and Lyman-$\alpha$ tracers~\cite{DESI:2025zpo}. These measurements, detailed in Table IV of Ref.~\cite{DESI:2025zgx}, span the {effective redshift range} $0.295 \leq z \leq 2.330$, divided into nine bins. BAO constraints are presented in terms of the transverse comoving distance $D_{\mathrm{M}}/r_{\mathrm{d}}$, the Hubble horizon $D_{\mathrm{H}}/r_{\mathrm{d}}$, and the angle-averaged distance $D_{\mathrm{V}}/r_{\mathrm{d}}$, all normalized to the comoving sound horizon at the drag epoch, $r_{\mathrm{d}}$. We also account for correlation structures through cross-correlation coefficients: $r_{V,M/H}$ and $r_{M,H}$, which capture the relationships between different BAO measurements. This dataset is referred to as \texttt{DESI-DR2}.
\\

\item \textit{Type Ia Supernovae} (\textbf{SN Ia}): Type Ia supernovae serve as standardizable candles, providing a fundamental probe of the Universe’s expansion history. Historically, SN Ia observations played a pivotal role in the discovery of the accelerating cosmic expansion~\cite{Filippenko:1998tv, perlmutter1999measurements}, reinforcing earlier theoretical and observational arguments for $\Lambda$-dominated cosmological models derived from large-scale structure data. 

In this work, we employ the following recent samples.  
\textbf{PantheonPlus and PantheonPlus\&SH0ES:} We incorporate the latest SN Ia distance modulus measurements from the PantheonPlus compilation~\cite{pantheonplus}, which includes 1701 light curves corresponding to 1550 distinct SN Ia events, spanning a redshift range of $0.01 < z < 2.26$. We refer to this dataset as \texttt{PP}. Additionally, we consider a variant calibrated using the most recent SH0ES Cepheid host distance anchors~\cite{Riess:2021jrx}, which determine the absolute magnitude of SN Ia independently of an external $H_0$ prior. This calibration yields a more robust and self-consistent determination, and we designate this dataset as \texttt{PP\&SH0ES}.

\end{itemize}

As a final note, we emphasize that throughout our analysis we adopt state-of-the-art assumptions for Big Bang Nucleosynthesis (BBN), sensitive to the constraints on the physical baryon density $\omega_{\rm b} = \Omega_{\rm b} h^2$. Specifically, the \texttt{BBN} likelihood that incorporates the most precise available measurements of primordial light element abundances: the helium mass fraction \( Y_P \), as determined in~\cite{Aver_2015}, and the deuterium-to-hydrogen ratio \( y_{\rm DP} = 10^5\, n_D / n_H \), from~\cite{Cooke_2018}. 

In what follows, we use the notation \texttt{Base} to denote the combination of SFRD and BBN constraints, i.e., \texttt{Base = SFRD + BBN}.

\begin{table*}
\centering
\caption{Best-fitting values of the baseline $\Lambda$CDM parameters obtained from different observational datasets. The quoted uncertainties correspond to the 68\% confidence level.}
\label{tab:parametros_LCDM}
\begin{tabular}{lcccc}
\hline
\textbf{Parameter} & \textbf{Base} & \textbf{+DESI-DR2} & \textbf{+PP} & \textbf{+PP\&SH0ES} \\
\hline
$H_0~[\mathrm{km\,s^{-1}\,Mpc^{-1}}]$ & $65\pm11$ & $68.28 \pm 0.18$ & $63^{+8}_{-10}$ & $73.6 \pm 1.0$ \\
$\Omega_{\mathrm{cdm}}$ &  $0.346^{+0.13}_{-0.091}$ & $0.2433 \pm 0.0061$ & $0.274 \pm 0.018$ & $0.2931 \pm 0.0200$ \\
$\alpha$ & $0.074^{+0.016}_{-0.034}$ & $0.094^{+0.018}_{-0.030}$ & $0.085^{+0.017}_{-0.030}$ & $0.084^{+0.017}_{-0.029}$ \\
$\beta$ & $ 0.102^{+0.010}_{-0.023}$ & $0.1183 \pm 0.0099$ & $0.109 \pm 0.010$ & $0.1096 \pm 0.0099$ \\
$\dot{\rho}_\star(z=0)~[\mathrm{M\odot\,yr^{-1}\,Mpc^{-3}}]$ & $0.0290^{+0.0030}_{-0.0041}$ & $0.0317^{+0.0025}_{-0.0029}$ & $0.0306^{+0.0024}_{-0.0029}$ & $0.0306^{+0.0024}_{-0.0029}$ \\
$\Omega_{\mathrm{m}}$ & $\Omega_m = 0.405^{+0.12}_{-0.099}$ & $0.2916 \pm 0.0058$ & $0.3340^{+0.0183}_{-0.0177}$ & $0.3341^{+0.0200}_{-0.0188}$ \\
\hline
\end{tabular}
\end{table*}

\begin{figure*}
    \centering
    \includegraphics[width=0.6\textwidth]{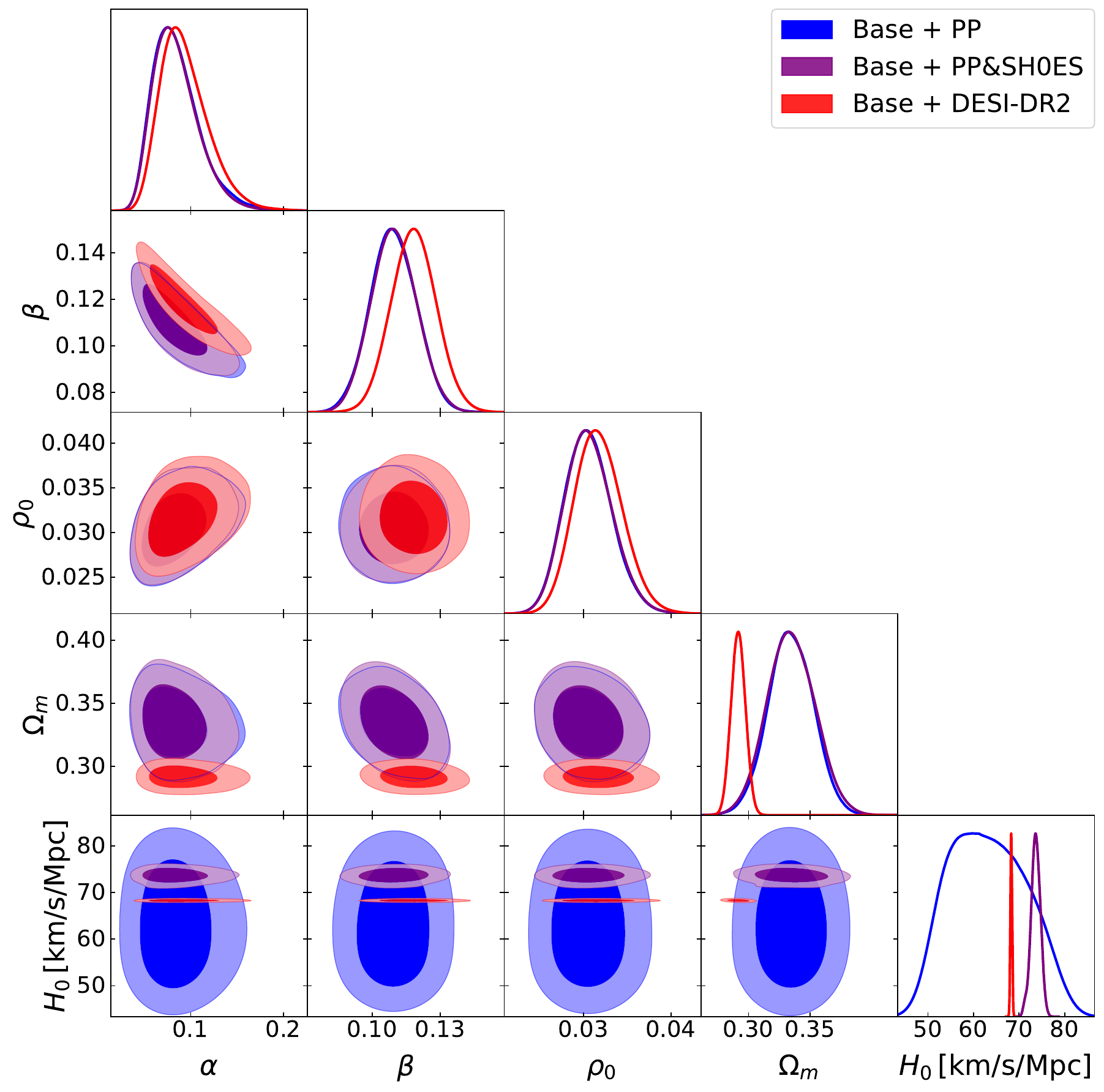}
    \caption{Triangle plot showing the joint posterior distributions of cosmological parameters ($\Omega_{\mathrm{m}}$, $H_0$) and astrophysical parameters ($\alpha$, $\beta$, and $\dot{\rho}_{\star,0}$). Constraints are derived from different combinations of likelihoods, as indicated in the legend.}
    \label{fig:triangle_comparacao}
\end{figure*}

\section{Results}
\label{results}

Table~\ref{tab:parametros_LCDM} summarizes our main statistical fits using different combinations of observational datasets. As expected, when considering only the observational probe defined as \texttt{Base}, i.e., constraints derived solely from SFRD + BBN measurements, the resulting parameter estimates exhibit large uncertainties \footnote{The analysis based solely on SFRD data (i.e., without BBN information) yields $H_0 = 75^{+23}_{-15}\ \mathrm{km\ s^{-1}\ Mpc^{-1}}$ and $\Omega_{\mathrm{cdm}} = 0.43^{+0.16}_{-0.24}$, both at 68\% CL.} This is a natural consequence of the limited constraining power of SFRD alone, which is sensitive to both cosmological and astrophysical parameters but suffers from degeneracies among them.

When additional cosmological probes are included—such as BAO from DESI or Type Ia supernovae—the constraints on the baseline $\Lambda$CDM parameters improve substantially. For example, the Hubble constant is tightly constrained by BAO measurements ($H_0 = 68.28 \pm 0.18~\mathrm{km\,s^{-1}\,Mpc^{-1}}$) and shifts toward higher values when PantheonPlus and SH0ES-calibrated supernovae are included, reflecting the well-known Hubble tension. The SFRD parameters ($\alpha$, $\beta$, and $\dot{\rho}_\star(z=0)$) remain broadly consistent across the different datasets, while their uncertainties are significantly reduced in the combined analyses, particularly when SFRD is jointly analyzed with BAO and SNIa data. In these joint constraints, the uncertainties decrease by approximately 30–50\% relative to the baseline case, without significant shifts in the central values. This indicates that astrophysical parameters associated with star formation can be more precisely determined when combined with robust cosmological probes. Overall, these improvements highlight the complementary nature of SFRD, BAO, and supernova data, leading to tighter and more reliable constraints on both cosmological and astrophysical parameters.

The cold dark matter density, $\Omega_{\mathrm{cdm}}$, exhibits a similar trend: SFRD alone allows only a broad range of values, whereas the inclusion of BAO and supernovae datasets substantially narrows the allowed parameter space. Notably, even though SFRD + BBN is less constraining than BAO or SNIa individually, it still provides complementary information on both cosmological and astrophysical parameters, highlighting the potential of SFRD as a complementary probe. 

It is instructive to compare our results with those reported by \cite{DESI:2025zgx}. Within the $\Lambda$CDM framework, the DESI collaboration, combining BAO and BBN data, finds $H_0 = 68.51 \pm 0.58~\mathrm{km,s^{-1},Mpc^{-1}}$. In contrast, our joint analysis (\texttt{Base + DESI-DR2}) yields $H_0 = 68.28 \pm 0.18~\mathrm{km,s^{-1},Mpc^{-1}}$. These results are in excellent agreement at the level of the central values, with a difference of $\Delta H_0 \simeq 0.23~\mathrm{km,s^{-1},Mpc^{-1}}$, well within the DESI uncertainty. Notably, our analysis achieves a significantly tighter constraint, reducing the uncertainty by approximately a factor of three. This improvement highlights the constraining power of our dataset combination, while remaining fully consistent with the standard $\Lambda$CDM scenario.
A similar level of agreement is found for the total matter density parameter, $\Omega_m = \Omega_{\mathrm{cdm}} + \Omega_b$. Our results are consistent with those of DESI within the quoted uncertainties, reinforcing the robustness of the $\Lambda$CDM framework across independent observational probes.

In comparison with the main analyses in \cite{pantheonplus,Riess:2021jrx}, we find that our baseline analysis (\texttt{Base + PP\&SH0ES}) yields $H_0 = 73.8 \pm 1.0~\mathrm{km,s^{-1},Mpc^{-1}}$, while the standard SH0ES team reports $H_0 = 73.3 \pm 1.1~\mathrm{km,s^{-1},Mpc^{-1}}$, including Cepheid distance calibrations. Therefore, the two measurements are fully consistent within their respective uncertainties and can be considered statistically equivalent. A similar conclusion can be drawn for the matter density parameter, $\Omega_m$. However, it is important to emphasize a key methodological difference between the analyses. In the SH0ES analysis using PP samples, $\Omega_m$ is effectively the only free cosmological parameter, with all other parameters fixed. In contrast, in our analysis we adopt a more general approach, allowing all relevant cosmological parameters in the baseline model to vary freely (e.g., the baryon density and $H_0$). As a consequence, our analysis naturally encompasses a larger parameter space and is subject to stronger parameter degeneracies. Despite this increased freedom, our results remain in remarkable agreement with those obtained from the PP and PP\&SH0ES analyses. This consistency reinforces the robustness of the inferred cosmological constraints across different modeling assumptions and parameter choices.

Overall, these results demonstrate that combining SFRD with established cosmological datasets not only helps break parameter degeneracies but also improves the statistical precision of astrophysical parameters associated with star formation. This underscores the dual role of SFRD as both a robust astrophysical diagnostic and a complementary cosmological probe.

Figure~\ref{fig:triangle_comparacao} shows the global parameter space of the model, including both cosmological parameters ($\Omega_{\mathrm{m}}$, $H_0$) and astrophysical parameters ($\alpha$, $\beta$, and $\dot{\rho}_\star(z=0)$). 

Given the high dimensionality of the parameter space (five parameters in the $\Lambda$CDM model) and the relatively weak constraining power of SFRD measurements alone, we adopt a strategy to reduce the effective number of free parameters in our analysis. Based on the correlations observed in Figure~\ref{fig:triangle_comparacao} (the main joint analysis \texttt{Base + DESI-DR2}), we parametrize the astrophysical quantities $\alpha$, $\beta$, and $\dot{\rho}_\star(z=0)$ as functions of the cosmological parameters $\Omega_{\mathrm{cdm}}$ and $h$. The goal is to retain only cosmological parameters in the MCMC sampling, effectively rewriting the astrophysical parameters ($\alpha$, $\beta$, and $\dot{\rho}_\star(z=0)$) as functions of cosmological quantities ($\Omega_{\mathrm{m}}$, $H_0$). Our methodology proceeds as follows:

These relations are obtained by performing functional fits directly to the posterior samples derived from the \texttt{Base + DESI-DR2} analysis. We explore a range of functional forms for this parametrization, including linear, quadratic, exponential, and logarithmic dependencies. For each astrophysical parameter, these candidate models are fitted to the sampled parameter space, and the optimal functional form is selected using the Bayesian Information Criterion (BIC) \cite{Liddle2007}. This approach enables us to identify the parametrizations that best capture the correlations present in the posterior distributions while avoiding overfitting. The numerical coefficients in the final expressions correspond to the best-fit values obtained from these functional regressions.

The resulting best-fit relations are then incorporated into a modified SFRD model, in which the astrophysical sector is fully mapped onto the cosmological parameter space. This procedure effectively reduces the dimensionality of the problem, mitigates degeneracies, and enables a more robust exploration of cosmological constraints driven by SFRD data.
\begin{equation}
\label{new_rho}
\dot{\rho}_\star(z) \;=\; 
\frac{\chi^{2}\,\dot{\rho}_{\star,0}^{\rm cosmo}}
{1 \;+\; \alpha^{\rm cosmo}\,(\chi-1)^{3}\,
\exp\!\big(\beta^{\rm cosmo}\,\chi^{7/4}\big)},
\end{equation}
where the parameters $\dot{\rho}_{\star,0}^{\rm cosmo}$, $\alpha^{\rm cosmo}$, and $\beta^{\rm cosmo}$ are now expressed as functions of $\Omega_{\mathrm{cdm}}$ and $h$, capturing the cosmology dependent behaviour of the star formation history in the form

\begin{align}
\alpha^{\rm cosmo} &= 0.091\,h - 0.3589\,\Omega_{\mathrm{cdm}} + 0.1209, \\
\beta^{\rm cosmo} &= 0.1647\,\exp\big(0.1132\,h - 1.703\,\Omega_{\mathrm{cdm}}\big), \\
\dot{\rho}_{\star,0}^{\rm cosmo} &= 0.0092\,h - 0.0328\,\Omega_{\mathrm{cdm}} + 0.0335.
\end{align}

The observed relationships among $\alpha$, $\beta$, $\dot{\rho}_\star(z=0)$, $h$, and $\Omega_{\mathrm{cdm}}$ can be understood in terms of the physical roles of these parameters.

An increase in $\Omega_{\mathrm{cdm}}$ enhances the abundance of massive halos, which in turn affects star formation. The parameter $\beta$ depends inversely on the variance $\sigma_4^2(z=0)$ of density fluctuations at halo scales with $T_{\rm vir} > 10^4~\mathrm{K}$. As higher $\Omega_{\mathrm{cdm}}$ boosts $\sigma_4$, $\beta$ decreases, reflecting the reduced suppression of star formation at high redshifts due to a greater number of efficient halos. Similarly, $\alpha$, which controls the early-time growth rate of the SFRD, decreases with increasing $\Omega_{\mathrm{cdm}}$, as the enhanced halo population diminishes the suppressive effect of $\alpha$. This interpretation relies on the strength of the negative correlation between $\Omega_{\mathrm{cdm}}$ and the parameters $\alpha$ and $\beta$.

The normalization, $\dot{\rho}_\star(z=0) \equiv \rho_0$, represents the present-day SFRD. Its dependence on $h$ and $\Omega_{\mathrm{cdm}}$ is indirect, since $\rho_0$ also incorporates factors such as star formation efficiency, gas availability, and feedback processes. Consequently, correlations of $\rho_0$ with cosmological parameters should be interpreted qualitatively rather than quantitatively.

Therefore, once the new equation~\ref{new_rho} is established, which allows a precise mapping between astrophysical and cosmological parameters, we can perform a new analysis in a reduced parameter space, focusing on parameter degeneracies and improving the precision of the estimates. Applying the method described above, we find that the inclusion of specific datasets significantly improves the constraints on key cosmological parameters relative to the baseline $\Lambda$CDM results reported in Table~\ref{tab:parametros_LCDM}. For the \texttt{Base + PP\&SH0ES} sample, we obtain $\Omega_{\mathrm{m}} = 0.329\pm 0.017
$ and $H_0 = 70.11\pm 0.24
~\mathrm{km\,s^{-1}\,Mpc^{-1}}$, while for the  \texttt{Base + PP} sample we find $\Omega_{\mathrm{m}} = 0.335^{+0.017}_{-0.015}$ and $H_0 = 66.97\pm 0.20~\mathrm{km\,s^{-1}\,Mpc^{-1}}$. In comparison, \texttt{Base + DESI-DR2} data yield $\Omega_{\mathrm{m}} = 0.2918 \pm 0.0063$ and $H_0 = 68.50 \pm 0.25~\mathrm{km\,s^{-1}\,Mpc^{-1}}$. 

\begin{table*}
\centering
\caption{Same as Table~\ref{tab:parametros_LCDM}, but for the $w$CDM model.}
\label{tab:parametros_wLCDM}
\begin{tabular}{lcccc}
\hline
\textbf{Parameter} & \textbf{Base} & \textbf{+DESI-DR2} & \textbf{+PP} & \textbf{+PP\&SH0ES} \\
\hline
$H_0~[\mathrm{km\,s^{-1}\,Mpc^{-1}}]$ & $69 \pm 17$ & $67.8 \pm 1.6$ & $66 \pm 15$ & $73.8 \pm 1.0$ \\
$\Omega_{\mathrm{cdm}}$ & $0.342^{+0.120}_{-0.088}$ & $0.2452 \pm 0.0066$ & $0.266^{+0.063}_{-0.056}$ & $0.302^{+0.060}_{-0.044}$ \\
$\alpha$ & $0.077^{+0.015}_{-0.037}$ & $0.094^{+0.019}_{-0.032}$ & $0.086^{+0.017}_{-0.031}$ & $0.083^{+0.017}_{-0.030}$ \\
$\beta$ & $0.102^{+0.011}_{-0.021}$ & $0.117 \pm 0.010$ & $0.112^{+0.011}_{-0.016}$ & $0.110^{+0.010}_{-0.016}$ \\
$\dot{\rho}_\star(z=0)~[\mathrm{M\odot\,yr^{-1}\,Mpc^{-3}}]$ & $0.0300^{+0.0032}_{-0.0048}$ & $0.0316^{+0.0026}_{-0.0030}$ & $0.0307^{+0.0024}_{-0.0029}$ & $0.0305^{+0.0025}_{-0.0029}$ \\
$\Omega_{\mathrm{m}}$ & $0.387^{+0.120}_{-0.088}$ & $0.2902 \pm 0.0066$ & $0.311^{+0.063}_{-0.056}$ & $0.347^{+0.060}_{-0.044}$ \\
$w$ & $-1.18^{+0.19}_{-0.34}$ & $-0.975 \pm 0.056$ & $-0.99^{+0.15}_{-0.12}$ & $-1.04^{+0.16}_{-0.14}$ \\
\hline
\end{tabular}
\end{table*}

\begin{figure*}
    \centering
    \includegraphics[width=0.6\textwidth]{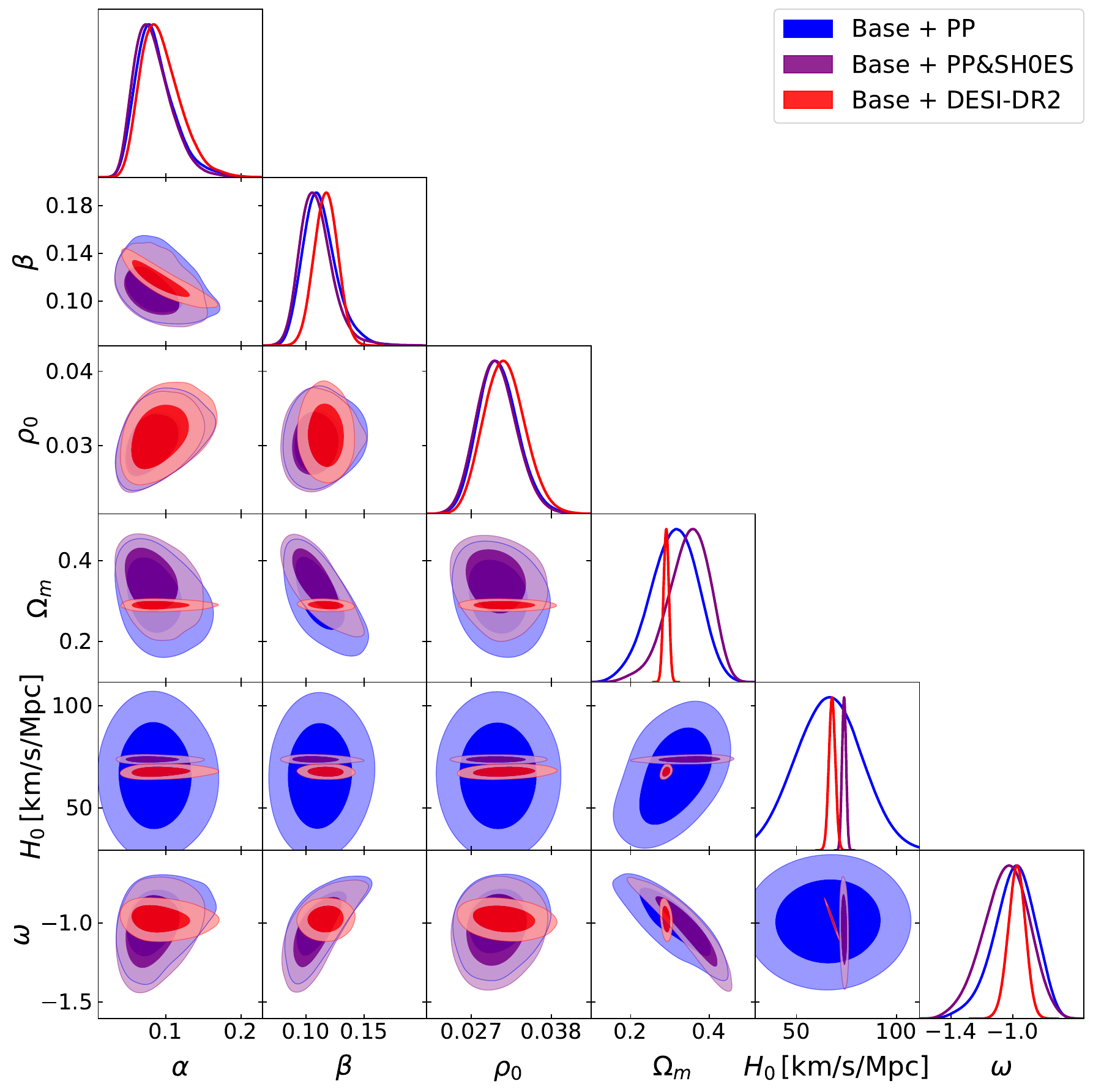}
    \caption{Triangle plot of the joint posterior distributions of cosmological parameters ($\Omega_{\mathrm{m}}$, $H_0$, $w$) and astrophysical parameters ($\alpha$, $\beta$, and $\dot{\rho}_{\star,0}$) for the $w$CDM model. Constraints are derived from different combinations of likelihoods, as indicated in the legend. This figure is analogous to Figure~\ref{fig:triangle_comparacao} for the $\Lambda$CDM case.}
    \label{fig:triangle_w0_comparacao}
\end{figure*}

Table~\ref{tab:parametros_wLCDM} summarizes the results obtained using the same data combinations and methodology as in the $\Lambda$CDM case, but applied to the $w$CDM scenario, where the dark energy equation-of-state parameter $w$ is allowed to vary. Introducing $w$ as a free parameter naturally increases degeneracies among the cosmological parameters, leading to broader uncertainties compared to the fixed-$w=-1$ $\Lambda$CDM case.

For the Hubble parameter, $H_0$, \texttt{DESI-DR2} continues to provide the tightest constraint ($H_0= 67.8 \pm 1.6~\mathrm{km\,s^{-1}\,Mpc^{-1}}$), highlighting its strong geometric constraining power. In contrast, \texttt{Base} alone allows a substantially broader range ($H_0 = 69 \pm 17~\mathrm{km\,s^{-1}\,Mpc^{-1}}$), reflecting the limited ability of high-redshift star formation data to break degeneracies between $H_0$ and $\Omega_{\mathrm{cdm}}$ when $w$ is free. The \texttt{PP} sample provides intermediate constraints ($H_0 = 66 \pm 15~\mathrm{km\,s^{-1}\,Mpc^{-1}}$), while \texttt{PP\&SH0ES} favors significantly higher values ($H_0 = 73.8 \pm 1.0~\mathrm{km\,s^{-1}\,Mpc^{-1}}$), consistent with the persistent Hubble tension between early- and late-universe measurements.

The cold dark matter density, $\Omega_{\mathrm{cdm}}$, follows a similar pattern. \texttt{DESI-DR2} tightly constrains $\Omega_{\mathrm{cdm}}$ around $0.2452 \pm 0.0066$, whereas \texttt{Base} allows a broad range ($0.342^{+0.120}_{-0.088}$). Base +  Pantheon datasets occupy intermediate values, with uncertainties larger than Base + BAO but smaller than Base alone. These trends illustrate how complementary probes differentially inform both the matter content and the expansion history of the Universe.

The dark energy equation-of-state parameter, $w$, is most precisely determined by \texttt{DESI-DR2} ($w = -0.975 \pm 0.056$), whereas \texttt{Base} provides only weak constraints ($w = -1.18^{+0.19}_{-0.34}$), highlighting that star formation data alone are insufficient to tightly constrain dark energy properties. \texttt{PP} and \texttt{PP\&SH0ES} yield results compatible with a cosmological constant ($w=-1$), but with significantly larger uncertainties than BAO, again demonstrating the need for geometric probes to reduce degeneracies in the $w$CDM model.

Astrophysical parameters associated with the SFRD, namely $\alpha$, $\beta$, and $\dot{\rho}_\star(z=0)$, remain broadly consistent with the $\Lambda$CDM results across all datasets. However, their uncertainties are slightly larger when $w$ is free, particularly in the SFRD + BBN combination, reflecting the additional degeneracy introduced by the free dark energy parameter. Despite this, the overall trends remain robust: $\alpha$ and $\beta$ continue to characterize the early- and late-time behavior of the cosmic star formation rate, while $\dot{\rho}_\star(z=0)$ sets the present-day normalization.

In summary, allowing $w$ to vary slightly broadens the allowed parameter space for both cosmological and astrophysical parameters, but does not qualitatively change the behavior observed in $\Lambda$CDM. The combination of multiple datasets SFRD, BAO, and Type Ia supernovae, remains essential for achieving tight and robust constraints in the $w$CDM framework, demonstrating the complementary nature of astrophysical and cosmological probes.

\begin{table}
\caption{Mean redshift of the peak of the cosmic star formation rate density for the $w$CDM and $\Lambda$CDM models, fitted to the indicated datasets. The peak redshift was computed by averaging the maxima of individual statistical realizations, and the uncertainties correspond to the range between the mean and the extrema of the distribution.}
\label{tab:redshift_picos}
\centering
\begin{tabular}{lcc}
\hline
\textbf{Model} & \textbf{Dataset} & \textbf{Peak Redshift, $z_{\rm peak}$} \\
\hline
$w$CDM & Base + PP\&SH0ES & $2.616^{+0.010}_{-0.192}$ \\
$w$CDM & Base + PP & $2.616^{+0.212}_{-0.192}$ \\
$w$CDM & Base + DESI-DR2 & $2.609^{+0.219}_{-0.185}$ \\
$\Lambda$CDM & Base + PP\&SH0ES & $2.595^{+0.119}_{-0.082}$ \\
$\Lambda$CDM & Base + PP & $2.553^{+0.161}_{-0.141}$ \\
$\Lambda$CDM & Base + DESI-DR2 & $2.600^{+0.114}_{-0.087}$ \\
\hline
\end{tabular}
\end{table}

\begin{figure*}
    \centering
    \includegraphics[width=0.45\textwidth]{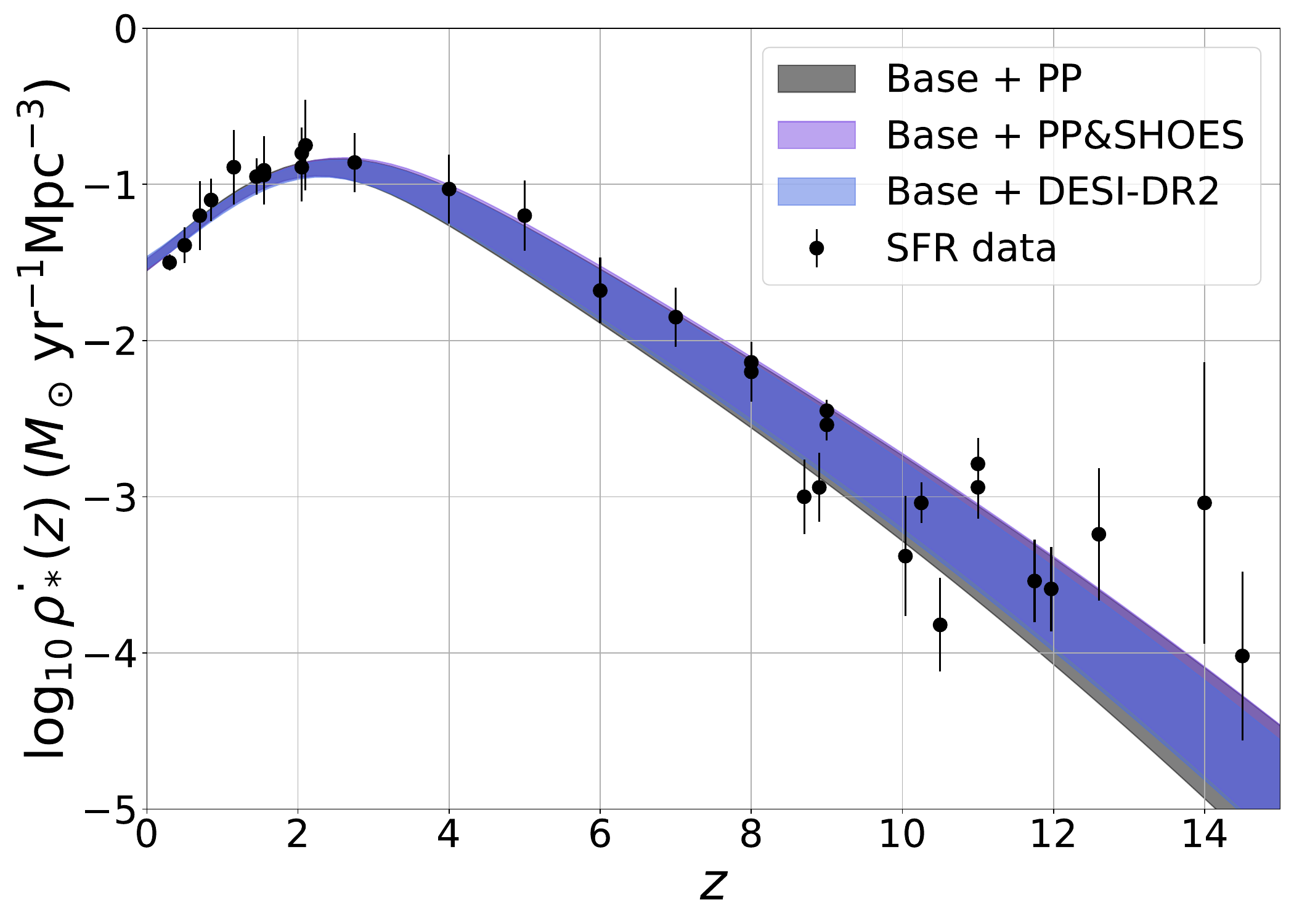}
    \hspace{0.04\textwidth}
    \includegraphics[width=0.45\textwidth]{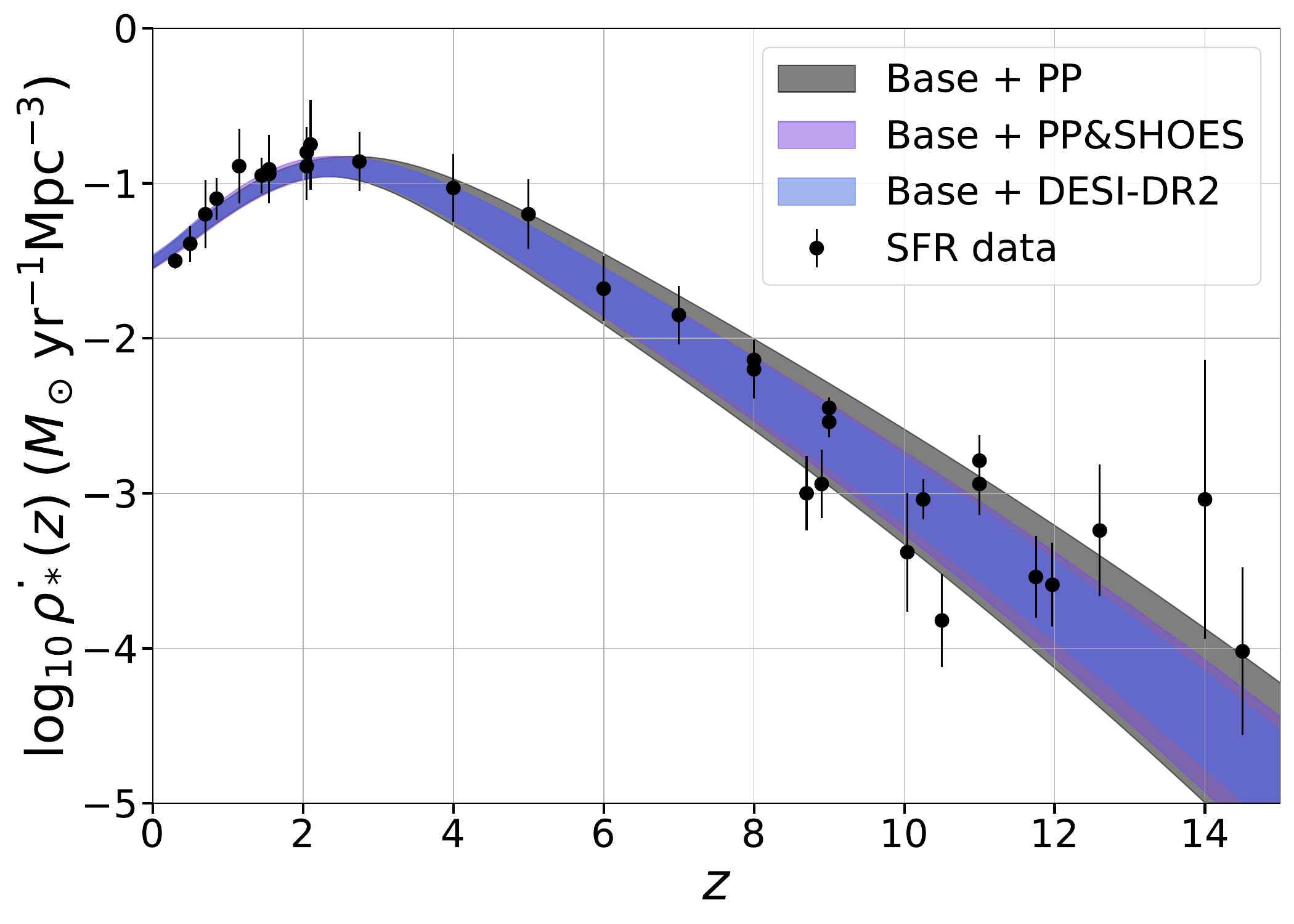}
    \caption{
        \textit{Left:} Uncertainty bands for the star formation rate under different cosmological datasets. 
        Observational data are shown in black, while the black, purple, and blue bands correspond to the 
        \texttt{Base + PP}, \texttt{Base + PP\&SH0ES}, and \texttt{Base + DESI-DR2} datasets, respectively. 
        \textit{Right:} Same as the left panel, but for the $w$CDM model.
    }
    \label{fig:sfr_curvas}
\end{figure*}

Table~\ref{tab:redshift_picos} summarizes the mean redshift of the peak of the SFRD for both the $w$CDM and $\Lambda$CDM models, fitted to the indicated datasets. The best-fit values and their associated uncertainties are inferred from the statistical reconstruction presented in Fig.~4, incorporating the best-fit estimates of all baseline parameters along with their corresponding uncertainties.

We observe that, for both cosmological models, the peak redshift values are remarkably consistent across the different datasets, including \texttt{PP\&SH0ES}, \texttt{PP}, and \texttt{DESI-DR2}. Differences between datasets are generally smaller than the 1$\sigma$ uncertainties, indicating that the determination of $z_{\rm peak}$ is robust against the choice of observational probe. In particular, \texttt{PP\&SH0ES} consistently produces slightly higher peak values compared to \texttt{PP}, while \texttt{DESI-DR2} yields intermediate values, reflecting the complementary constraining power of geometric and astrophysical datasets.

Comparing the two cosmological scenarios, $w$CDM generally predicts marginally higher peak redshifts than $\Lambda$CDM, though these differences remain within the statistical uncertainties. For example, using \texttt{PP\&SH0ES}, we find $z_{\rm peak} = 2.616^{+0.010}_{-0.192}$ for $w$CDM, versus $z_{\rm peak} = 2.595^{+0.119}_{-0.082}$ for $\Lambda$CDM. This systematic shift is consistent with the slightly different expansion histories in $w$CDM, where a free dark energy equation-of-state can modify the growth and assembly of halos at early times, subtly affecting the timing of peak star formation.

Overall, these results demonstrate that the global SFRD peak is robust with respect to both the choice of cosmological model and the dataset. Across all models and probes considered, cosmic star formation activity reached its maximum around $z \sim 2.6$, in agreement with some previous studies of the cosmic star formation history \cite{Liu_2025, Novak_2017, 2014A&A...571A..30P}. 
Therefore, at least within the context considered here, the peak of Cosmic Noon appears to be insensitive to the nature of dark energy, for the two models analyzed.

Figure~\ref{fig:sfr_curvas} presents a statistical reconstruction of the cosmic star formation rate density (SFRD) based on the datasets updated in this work. The left panel corresponds to the reconstruction under the $\Lambda$CDM model, while the right panel shows the results for the $w$CDM scenario. From these plots, it is evident that the framework adopted here is capable of simultaneously fitting both cosmological and astrophysical parameters in light of the SFRD measurements. In particular, the reconstructed curves capture the overall shape of the star formation history, including the rise at high redshift, the peak around $z \sim 2.6$, and the subsequent decline toward the present day. The comparison between the two cosmological models highlights subtle differences in the SFRD evolution, reflecting the impact of varying the dark energy equation-of-state parameter on the growth of structure and the efficiency of star formation in halos of different masses.
It is important to note that measurements at high redshift ($z \gtrsim 8$) are still subject to significant uncertainties and increased dispersion, reflecting both observational limitations and modeling fit challenges from a observational point of view. Consequently, the acquisition of additional data and the development of more robust and homogeneous compilations of SFRD measurements in this regime will be crucial. Such improvements may open new avenues for constraining cosmological parameters.

\section{Final Remarks}
\label{final}

In this work, we have explored the potential of the cosmic star formation rate density (SFRD) as a complementary cosmological probe, emphasizing its capability to simultaneously constrain both astrophysical and cosmological parameters. Utilizing datasets spanning a wide redshift range, $z \in [0, 15]$, we derived constraints within the $\Lambda$CDM and $w$CDM frameworks, considering various combinations with BAO measurements and Type Ia supernovae (PantheonPlus and Pantheon+SH0ES).

Our analysis shows that \texttt{Base} probe (SFRD + BBN), provides relatively weak constraints on cosmological parameters, primarily due to degeneracies with astrophysical factors such as star formation efficiency and feedback processes. Nonetheless, the \texttt{Base} results remain broadly consistent with other datasets. When combined with BAO and SNIa data, these degeneracies are effectively broken, leading to substantial improvements in the precision of key cosmological parameters, including $H_0$ and $\Omega_{\rm cdm}$, as well as the astrophysical SFRD parameters, i.e.,$\alpha$, $\beta$, $\dot{\rho}_\star(z=0)$. For example, in the $\Lambda$CDM framework, the joint \texttt{Base} + \texttt{DESI-DR2} analysis constrains the Hubble parameter to $H_0 = 68.28 \pm 0.18$ km\,s$^{-1}$\,Mpc$^{-1}$, compared to the much broader range obtained with \texttt{Base} alone, $H_0 = 65\pm11$ km\,s$^{-1}$\,Mpc$^{-1}$.

We also investigated the parametric correlations between astrophysical and cosmological quantities. By expressing the SFRD parameters ($\alpha$, $\beta$, $\dot{\rho}_\star(z=0)$) as functions of cosmological parameters ($\Omega_{\rm cdm}$, $h$), we demonstrated how the physical properties of star-forming halos are intimately connected to the underlying cosmology. This approach allows for a more unified understanding of the cosmic star formation history and its dependence on the matter content and expansion rate of the Universe.

Extending the analysis to the $w$CDM model, we observe similar trends, albeit with slightly larger uncertainties due to the freedom in the dark energy equation-of-state parameter, $w$. The peak of cosmic star formation remains robust across both cosmological models and datasets, occurring around $z_{\rm peak} \sim 2.6$, with only minor variations between $\Lambda$CDM and $w$CDM. Moreover, the SFRD parameters remain remarkably stable, confirming the reliability of the astrophysical parametrization even when dark energy is allowed to deviate from a cosmological constant.

Overall, our results illustrate that the SFRD serves as a probe at the interface between cosmology and galaxy formation. When combined with established geometrical probes, it provides complementary constraints, reduces parameter degeneracies, and enhances the statistical precision of both cosmological and astrophysical quantities. On the other hand, cosmological observations up to redshift $z \approx 2$ are currently measured with remarkable precision, particularly through well-established probes such as SNIa and BAO. However, beyond this redshift range, traditional cosmological probes become increasingly sparse and less constraining. In this context, future measurements of the SFRD at higher redshifts, specially those enabled JWST, offer a promising avenue to extend our observational reach into previously unexplored epochs of cosmic history. By probing the Universe at redshifts well beyond the limits of SNIa and BAO, these observations can provide complementary and independent constraints on cosmological models. In particular, improved high-redshift SFRD data have the potential to enhance its role as a sensitive cosmological diagnostic, shedding light on the interplay between cosmological parameters, baryonic physics, and the processes governing early star formation.

\section*{Acknowledgements}
\noindent We thank the referee for the careful reading of the manuscript and for several constructive comments and suggestions, which helped us to improve the clarity and presentation of the manuscript. The authors thank Rogério Riffel for valuable discussions on observational techniques in cosmic star formation. M. M received support from the FAPERGS scholarship. R.C.N. thanks the financial support from the Conselho Nacional de Desenvolvimento Científico e Tecnologico (CNPq, National Council for Scientific and Technological Development) under the project No. 304306/2022-3, and the Fundação de Amparo à Pesquisa do Estado do RS (FAPERGS, Research Support Foundation of the State of RS) for partial financial support under the project No. 23/2551-0000848-3.

\section*{Data Availability}
The datasets and products used in this research, including Boltzmann codes and likelihoods, will be made available upon reasonable request to the corresponding author following the publication of this article.

\appendix
\section{Fitting Star Formation Rate Data with Additional Parametric Models}

A relevant question is whether our inferred value of the peak redshift, $z_{\mathrm{peak}}$, depends on the specific functional form adopted to describe the star formation rate density (SFRD). In particular, one may ask whether similar results would be obtained if, instead of Eq.~\ref{main_SFR}, we adopted widely used parametrizations such as those proposed in  Madau \& Dickinson (2014) \cite{Madau:2014bja}

\begin{equation}
\label{madau_dickinson }
\dot{\rho_{\star}}(z) = \dot{\rho_{\star,0}}\frac{(1+z)^{a}}{1 + [(1+z)/c]^{b}}.
\end{equation}

Alternatively, as discussed in, e.g., \cite{Harikane_2022}, the Madau \& Dickinson parametrization fails to accurately describe SFRD data at high redshift (e.g., $z \gtrsim 5$; see Figs.~23 and 24 therein). This limitation arises because it was calibrated using datasets with large uncertainties and sparse observational coverage in this regime, and is typically constrained to $z \lesssim 8$. As a result, the fit is not well anchored at high redshifts. Recent observations, particularly from the JWST, further highlight this issue, indicating galaxy abundances and properties that are not well captured by this parametrization. Therefore, in addition to the Madau \& Dickinson model, we also consider an alternative parametrization proposed in \cite{Harikane_2022}, which is designed to fit data up to $z > 10$, and is given by

\begin{equation}
\label{Harikane}
\dot{\rho_{\star}}(z) =
\frac{1}{
a(1+z)^b + c\,10^{d(1+z)} + e\,10^{f(1+z)+g}}.
\end{equation}

We perform an analysis analogous to that applied to Eq.~(\ref{main_SFR}), now considering both fitting approaches described above.



\begin{table*}[t]
\centering
\caption{Constraints obtained using SFRD data only for the Madau \& Dickinson \cite{Madau:2014bja} and Harikane et al. \cite{Harikane_2022} parametrizations.}
\begin{tabular}{|l|c|c|}
\hline\hline
\textbf{Parameter} & \textbf{Madau \& Dickinson} & \textbf{Harikane et al.} \\ 
\hline
$a$ & $1.65 \pm 0.26$ & $59^{+10}_{-10}$ \\
$b$ & $8.66^{+0.59}_{-0.49}$ & $-2.46^{+0.48}_{-0.39}$ \\
$c$ & $5.11^{+0.33}_{-0.28}$ & $0.382^{+0.073}_{-0.21}$ \\
$d$ & --- & $0.293^{+0.023}_{-0.017}$ \\
$e$ & --- & $4.7^{+2.2}_{-3.8}$ \\
$f$ & --- & $0.33^{+0.13}_{-0.27}$ \\
$g$ & --- & $-6.2^{+2.0}_{-3.3}$ \\
$\dot{\rho}_{\star,0}\ [\mathrm{M_\odot\ yr^{-1}\ Mpc^{-3}}]$ & $0.0228^{+0.0028}_{-0.0042}$ & --- \\
\hline\hline
\end{tabular}
\label{tab:combined_fits}
\end{table*}

\begin{figure}[h]
\centering
\includegraphics[width=0.5\textwidth]{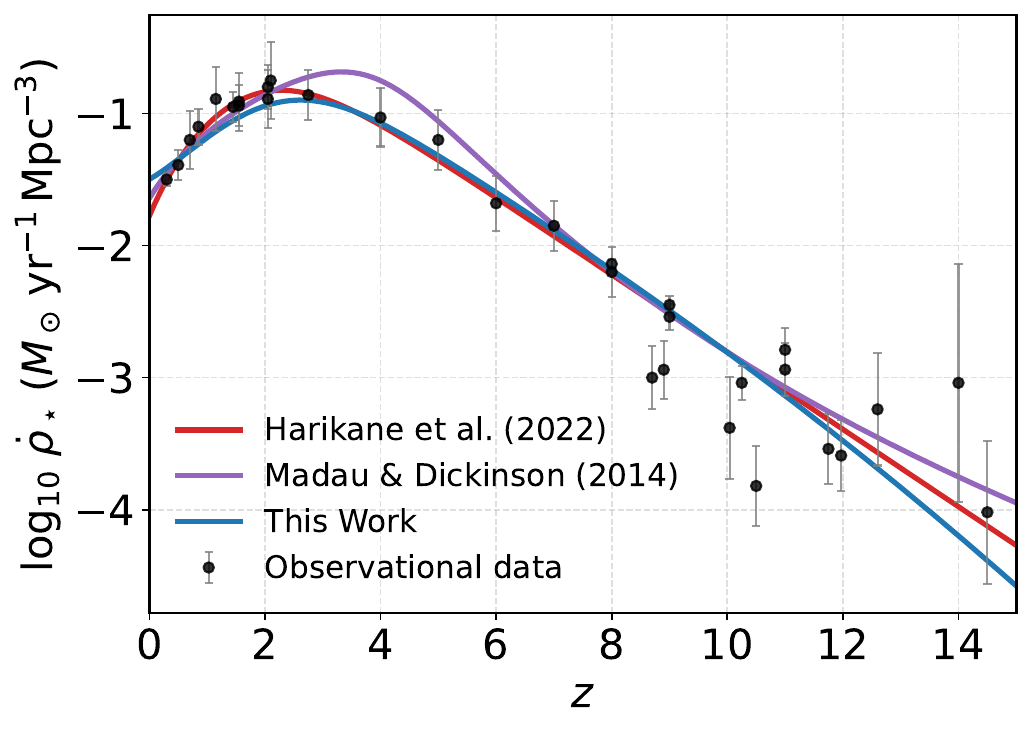}
\caption{Star formation rate density (SFRD) as a function of redshift. The red and purple curves correspond to the parametrizations of Harikane et al. (2022) and Madau \& Dickinson (2014), respectively, both fitted to the SFRD data used in this work. The blue curve (``This Work'') shows the fit obtained from Eq.~(\ref{main_SFR}), with parameters constrained using the combined \texttt{Base + DESI-DR2} datasets. Black points represent the observational data.}
\label{fig:sfr_md}
\end{figure}

Table \ref{tab:combined_fits} summarizes the statistical constraints obtained for the Madau \& Dickinson and Harikane et al.\ parameterizations. Figure \ref{fig:sfr_md} shows the corresponding fits to the SFRD data, where, in each case, the free parameters are fixed to their respective best-fit values.

Regarding the peak of the cosmic star formation rate density, $z_{\mathrm{peak}}$, we find $z_{\mathrm{peak}} = 3.326^{+0.086}_{-0.084}$ for the Madau \& Dickinson parameterization and $z_{\mathrm{peak}} = 2.295^{+0.166}_{-0.134}$ for the Harikane et al.\ parameterization. This discrepancy reflects the different functional forms and underlying assumptions of the two models, particularly in their treatment of the high-redshift regime. In the Madau \& Dickinson case, the peak is shifted to higher redshifts due to its more restrictive functional form and limited flexibility beyond $z \gtrsim 4$. In contrast, the Harikane et al.\ parameterization, which is designed to better accommodate recent high-redshift observations, yields a peak at lower redshift and a broader evolution of the SFRD. These differences highlight the importance of the chosen parametrization, especially in light of increasingly precise high-redshift data.

We now compare these fits with our main results, from which we obtain a constraint of $z_{\mathrm{peak}} = 2.600^{+0.114}_{-0.087}$, derived within the $\Lambda$CDM framework using the combined \texttt{Base + DESI-DR2} dataset. This value lies between the estimates obtained from the Madau \& Dickinson and Harikane et al.\ parameterizations, highlighting the sensitivity of $z_{\mathrm{peak}}$ to the assumed functional form of the SFRD and to the choice of data samples. Overall, this comparison underscores the importance of adopting parametrizations that can reliably capture the high-redshift regime when deriving robust estimates of key quantities such as $z_{\mathrm{peak}}$.

It is important to emphasize that parameterizations such as those of Madau \& Dickinson and Harikane et al.\ are essentially polynomial fits to the data. Their functional forms are not directly derived from any astrophysical modeling principles, and their parameters do not carry a clear physical interpretation. As a consequence, these models do not explicitly encode the underlying astrophysical processes driving star formation, nor do they provide any direct constraints on cosmological parameters.

Perhaps more importantly for the present discussion, although different parametrizations can lead to noticeably different values of $z_{\mathrm{peak}}$, their inferred trends appear to be largely insensitive to the underlying cosmological model, at least within the baseline scenarios explored in this work. This suggests that the dominant source of variation in $z_{\mathrm{peak}}$ arises from the adopted SFRD functional form rather than from the background cosmology itself. Therefore, caution is required when interpreting $z_{\mathrm{peak}}$ as a cosmological probe, as its value may be significantly driven by modeling choices rather than by fundamental physics.

\bibliography{main}
\end{document}